\documentclass{article}
\usepackage{icmc2021template}
\usepackage{times}
\usepackage{ifpdf}
\usepackage{soul}
\usepackage{float}
\usepackage{amsmath}
\usepackage[utf8]{inputenc}
\usepackage[english]{babel}
\usepackage{xcolor}
\usepackage[english]{babel}
\def\papertitle{Sketching sounds: an exploratory study on sound-shape associations}
\def\firstauthor{Sebastian Löbbers}
\def\secondauthor{Mathieu Barthet}
\def\thirdauthor{György Fazekas}

\newif\ifpdf
\ifx\pdfoutput\relax
\else
   \ifcase\pdfoutput
      \pdffalse
   \else
      \pdftrue
  \fi
\fi

\ifpdf % compiling with pdflatex
  \usepackage[pdftex,
    pdftitle={\papertitle},
    pdfauthor={\firstauthor, \secondauthor},
    bookmarksnumbered, % use section numbers with bookmarks
    pdfstartview=XYZ % start with zoom=100% instead of full screen; 
   ]{hyperref}
  \usepackage[pdftex]{graphicx}
  \graphicspath{{./figures/}}
  \DeclareGraphicsExtensions{.pdf,.jpeg,.png}

  \usepackage[figure,table]{hypcap}

\else % compiling with latex
  \usepackage[dvips,
    bookmarksnumbered, % use section numbers with bookmarks
    pdfstartview=XYZ % start with zoom=100% instead of full screen
  ]{hyperref}  % hyperrefs are active in the pdf file after conversion

  \usepackage[dvips]{epsfig,graphicx}
  \graphicspath{{./figures/}}
  \DeclareGraphicsExtensions{.eps}

  \usepackage[figure,table]{hypcap}
\fi

\hypersetup{
    colorlinks,%
    citecolor=black,%
    filecolor=black,%
    linkcolor=black,%
    urlcolor=black
}

\title{\papertitle}

 \threeauthors
   {0.5in}
   {\firstauthor} {Centre for Digital Music \\ Queen Mary University of London \\ % 
   {\tt \href{mailto:s.lobbers@qmul.ac.uk}{s.lobbers@qmul.ac.uk}}}
   {\secondauthor} {Centre for Digital Music \\ Queen Mary University of London \\ %
     {\tt \href{mailto:m.barthet@qmul.ac.uk}{m.barthet@qmul.ac.uk}}}
    {\thirdauthor} {Centre for Digital Music \\ Queen Mary University of London \\ %
     {\tt \href{mailto:g.fazekas@qmul.ac.uk}{g.fazekas@qmul.ac.uk}}}

\begin{document}
\capstartfalse
\maketitle
\capstarttrue
\begin{abstract}
Sound synthesiser controls typically correspond to technical parameters of signal processing algorithms rather than intuitive sound descriptors that relate to human perception of sound. This makes it difficult to realise sound ideas in a straightforward way. Cross-modal mappings, for example between gestures and sound, have been suggested as a more intuitive control mechanism. A large body of research shows consistency in human associations between sounds and shapes. However, the use of drawings to drive sound synthesis has not been explored to its full extent. This paper presents an exploratory study that asked participants to sketch visual imagery of sounds with a monochromatic digital drawing interface, with the aim to identify different representational approaches and determine whether timbral sound characteristics can be communicated reliably through visual sketches. Results imply that the development of a synthesiser exploiting sound-shape associations is feasible, but a larger and more focused dataset is needed in followup studies. 

\end{abstract}

\section{Introduction}\label{sec:introduction}
Timbre is an increasingly significant component of modern music production, often serving as a distinguishing characteristic of an artist's style or genre. Blake~\cite{blakeTimbreDifferentiationIndie2012} provides an example by describing how rock bands can be divided into sub-genres by analysing timbre rather than chord progression, referencing the groups My Bloody Valentine and U2. Despite humans' capability of perceiving very subtle differences in timbre~\cite{pengWhyCanYou2018} it is still poorly understood by researchers~\cite{saitisTimbreSemanticsLens2018,siedenburgComparisonApproachesTimbre2016,mcadamsPerceptionMusicalTimbre2016}. Crafting the ``right" sound is an important part of electronic music production, but synthesiser parameters typically correspond to a technical function related to signal processing, rather than a concept related to human perception, making it difficult to realise sound ideas in a straightforward way. The aim of this research is to develop a sketch-based sound synthesiser that takes simple drawings as input to provide a simple, intuitive control informed by cross-modal associations between sounds and shapes. The development will be informed by the results of a study, presented in this paper, that explores how participants represent timbre with a digital drawing interface. This section introduces relevant research into sound-shape associations and presents related work about sound synthesis control. Sections \ref{sec:methods}, \ref{sec:analysis} and \ref{sec:results} describe the design, analysis and results of the study. The discussion in Section \ref{sec:discussion} is followed by a conclusion in Section \ref{sec:conclusion}.

\subsection{Background on Sound-Shape Associations}
\label{sec:background}
Strong evidence suggests that a majority of people think of sound in a visual way to some extent that references colour, brightness, shapes and contour~\cite{martinoSynesthesiaStrongWeak2001}. In the 1920s, K{\"o}hler discovered that humans associate the made-up word \textit{takete} with sharp, jagged shapes, and \textit{maluma} with soft, round shapes~\cite{kohler1929gestalt}. Similar findings were made with the words \textit{boubou} and \textit{kiki}~\cite{ramachandranSynaesthesiaWindowPerception} and generally among all phonemes~\cite{nielsenParsingRoleConsonants2013}. This effect has been observed across cultures~\cite{davisFitnessNamesDrawings1961,taylorPhoneticSymbolismFour1962,bremnerBoubaKikiNamibia2013}, age groups including toddlers~\cite{maurerShapeBoubasSound2006} and, to some extent, with the visually impaired~\cite{bottiniSoundSymbolismSighted2019}. Adeli et al.~\cite{adeliAudiovisualCorrespondenceMusical2014} and Grill et al.~\cite{grillVisualizationPerceptualQualities2012} found similar associations between shapes and musical instruments or abstract sonic textures respectively. Focusing on pitch, loudness and tempo, K{\"u}ssner et al.~\cite{kussnerMusiciansAreMore2014} asked participants to draw their representations of sound rather than selecting existing shapes. They found that representations are not only influenced by musical structure, but also by a participant's music proficiency. Engeln and Groh~\cite{engelnCoHEARenceAudibleShapes2020} loosely classified drawings of sounds that were re-synthesised from spectrograms into real-world associations like scenes, actions or emotions, abstract shapes and structures or references to audio visualisations like waveforms and amplitude envelopes. 

\subsection{Related Work}
\label{sec:related_work}
A variety of strategies has been proposed to improve the intuitiveness of sound synthesisers. Low-level synthesis parameters can be mapped to a smaller number of high-level descriptors that either correspond to concepts of human perception as implemented in the \textit{FeatSynth} framework~\cite{hoffmanFeatureBasedSynthesisMapping} or describe the actions and objects that produce a sound, as seen with the impact sound synthesiser developed at the PRISM laboratory~\cite{bourachotPerceptionObjectAttributes2019,aramakiControllingPerceivedMaterial2011}. Further, a different input modality like gestures~\cite{eslingFlowSynthSimplifyingComplex2020} or voice~\cite{ekmanUsingVocalSketching2010} allows for a more intuitive control, potentially affording the simultaneous manipulation of multiple parameters. This can be particularly helpful for the exploration and drafting of sound ideas. Timbre visualisations, as mentioned in section \ref{sec:background}, are more frequently used in a sound retrieval context~\cite{richanProposalEvaluationNew2020}, but have also been adapted for synthesis, for example by \textit{Sound Mosaic}~\cite{giannakisComparativeEvaluationAuditoryvisual2006} which allows users to manipulate shapes to drive sound synthesis. Knees and Andersen~\cite{kneesSearchingAudioSketching2016} explored how drawings could be used for sound retrieval with a non-functioning prototype, an approach more commonly found in image search applications~\cite{sousaSketchbasedRetrievalDrawings2010,zhangSketchBasedImageRetrieval2016}. A major challenge of developing a sketch-based sound synthesiser is to find meaningful mappings between visual features and synthesis parameters.

\section{Methods and Material}
\label{sec:methods}
This section describes the design of an exploratory study that investigates how participants represent sound stimuli intuitively through free-form sketching with a digital drawing interface. The following hypotheses were put forward:
\begin{itemize}
    \item There will be some level of agreement between participants on how to sketch a sound.
    \item Correlations between quantitative sketch and audio features will align with the sound-shape associations described in Section \ref{sec:background}.
    \item A participant's music proficiency and the sound type will have an influence on the representational approach, but overall abstract sketches will be produced more frequently than depictions of real-world associations.  
\end{itemize}
%, Queen Mary University of London 
\subsection{Participants}
\label{sec:participants}
Twenty-eight participants were recruited through mailing lists and in person at the School of Electronic Engineering and Computer Science at Queen Mary University of London. This group was divided equally by gender (14 female, 14 male), 25 were adults below the age of 33 (three between 34 and 49), 22 had a Western background (16 Europe, 4 North America, 2 South America) and 5 an Eastern background (4 China, 1 India) with one participant preferring not to disclose this information. As described in Section \ref{sec:procedure}, participants were divided by music proficiency resulting in 14 musicians and 14 non-musicians.

\subsection{Stimuli} 
\label{sec:stimuli}
A total of ten timbrally dissimilar sounds were selected following the research of Adeli et al.~\cite{adeliAudiovisualCorrespondenceMusical2014} and Grill et al.~\cite{grillVisualizationPerceptualQualities2012} ranging from musical instruments (\textit{Piano}, \textit{Strings}, \textit{Electric Guitar}) and environmental sounds (\textit{Impact}) to synthesised pads (\textit{Telephonic}, \textit{Subbass}) and abstract textures (\textit{Noise}, \textit{String Grains}, \textit{Crackles}, \textit{Processed Guitar}).\footnote{All sounds can be accessed online together with the sketches drawn by participants during the experiment \url{https://bit.ly/3ta6crU}.} All sound stimuli are monophonic, normalised for equal loudness, pitched to the MIDI note C3 and last eight seconds including trailing silence to mark a clear endpoint during looped playback. The perceived base frequency may vary due to prominent harmonics. 
 
\subsection{Apparatus}
The drawing interface was implemented in \textit{p5.js} and runs in a web-browser. White strokes can be drawn on a 750x750 pixel canvas with a black background that separates it from the rest of the page. Stroke colour or width cannot be changed and sketches cannot be modified or erased. This design was chosen to encourage participants to follow their intuition and focus on shape rather than visual texture and colour. Clicking and dragging the mouse cursor starts a sketch and the timestamped cursor position is recorded consecutively while sketching. The study was conducted in person using the trackpad on a 15" MacBook Pro and a pair of Beyerdynamic DT 770 Pro headphones in calm, indoor locations.

\subsection{Procedure}
\label{sec:procedure}
Participants first completed a questionnaire that included an excerpt of the Gold MSI framework~\cite{mullensiefenGoldsmithsMusicalSophistication} and was used to determine their music proficiency.\footnote{A participant was categorised as a musician if they scored above average and reported involvement in musical activity.} Participants were asked to familiarise themselves with the drawing interface without audio before they were presented with the sound stimuli. The study intended to encourage a spontaneous response, therefore no information about the range of sounds was provided and participants were instructed to sketch what they believed to represent each sound stimulus the best. Looped playback started automatically with the option to pause/resume. Each sound was played twice in a randomised order resulting in a total of twenty sketches per participant. After completion, a short semi-structured interview was conducted and audio recorded asking participants how they approached the task and whether they found it difficult. No time limit was given and the study typically took twenty to thirty minutes to complete. The study setup can be accessed online.\footnote{Study setup: \url{https://bit.ly/3j3FkVO}}

\begin{table*}[t]
 \begin{center}
 \begin{tabular}{|p{0.07\linewidth}|p{0.07\linewidth}||p{0.07\linewidth}|p{0.07\linewidth}||p{0.07\linewidth}|p{0.07\linewidth}||p{0.07\linewidth}|p{0.07\linewidth}||p{0.07\linewidth}|p{0.07\linewidth}|}
  
    \multicolumn{2}{l}{\textbf{Grains}} 
    & 
    \multicolumn{2}{l}{\textbf{Lines}}
    & 
    \multicolumn{2}{l}{\textbf{Object/Scenes}} 
    & 
    \multicolumn{2}{l}{\textbf{Chaotic/Jagged}} 
    & 
    \multicolumn{2}{l}{\textbf{Radiating}} 
    \\
    \hline
    \multicolumn{2}{|p{0.14\linewidth}||}{{\tiny \textit{Small, repeated, grainy, spots, multiple components, layers, abstract, distinct}}} 
    &
    \multicolumn{2}{|p{0.14\linewidth}||}{{\tiny \textit{round, soft, continuous, jagged, irregular, simple, single, lines}}}
    & 
    \multicolumn{2}{|p{0.14\linewidth}||}{{\tiny \textit{real-life objects, environment, actions or feelings, abstract structures}}} 
    & 
    \multicolumn{2}{|p{0.14\linewidth}||}{{\tiny \textit{chaotic, intense, jagged, multiple layers, single objects}}} 
    & 
    \multicolumn{2}{|p{0.14\linewidth}||}{{\tiny \textit{round, circular, spiral, sharp, shaking, distinct objects, radiating, natural}}}
    \\
  \hline
  \hline
    \includegraphics[width=1.05\linewidth]{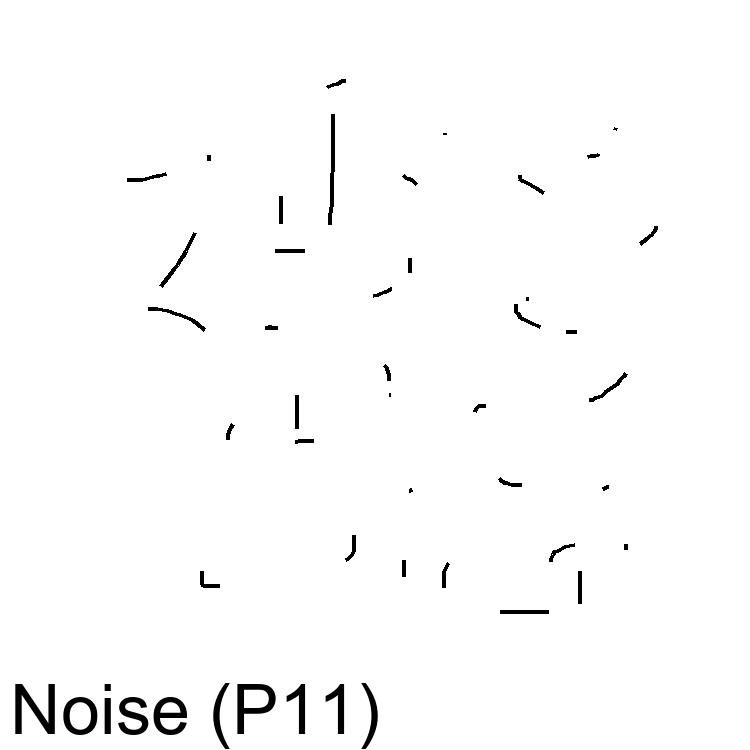} 
    &
    \includegraphics[width=1.05\linewidth]{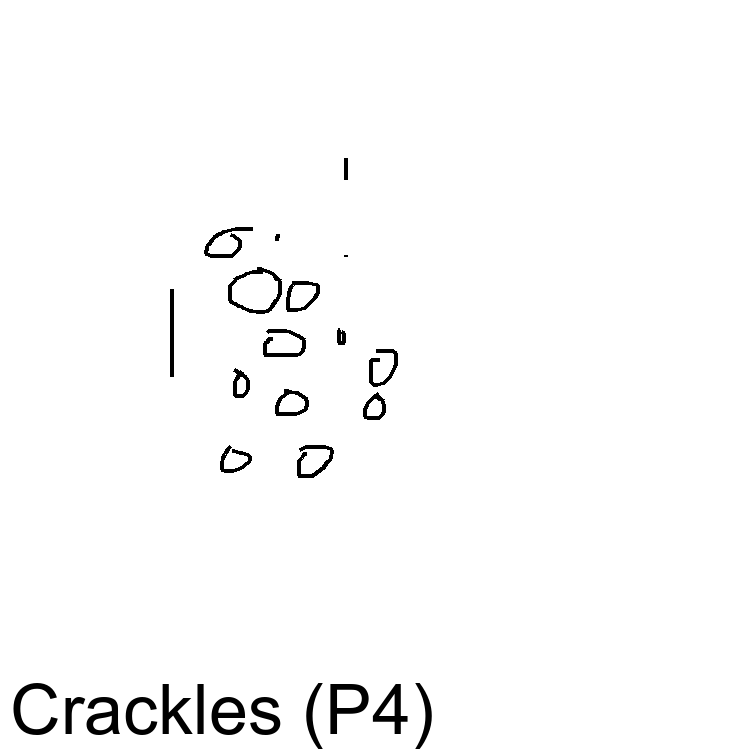} 
    &
    \includegraphics[width=1.05\linewidth]{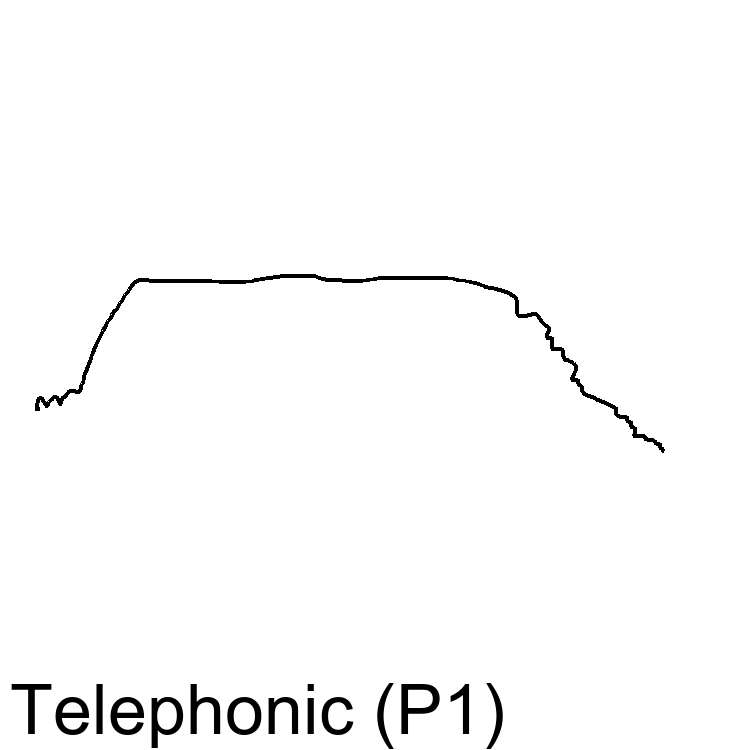} 
    &
    \includegraphics[width=1.05\linewidth]{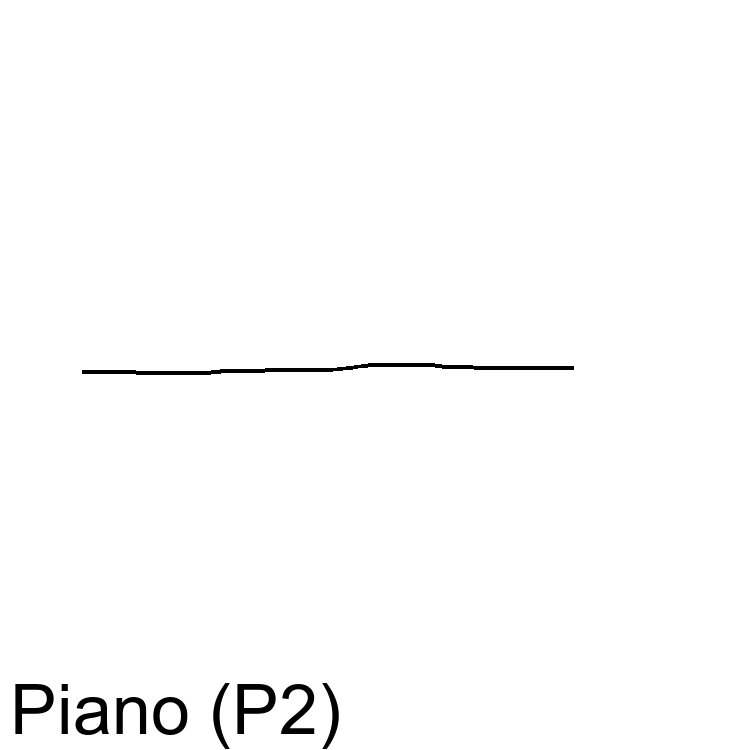} 
    &
    \includegraphics[width=1.05\linewidth]{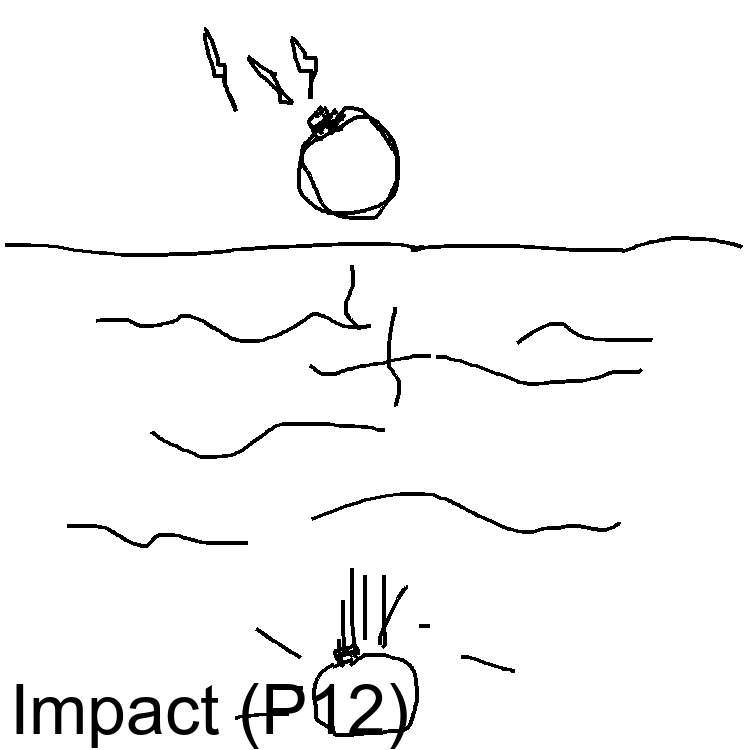} 
    & 
    \includegraphics[width=1.05\linewidth]{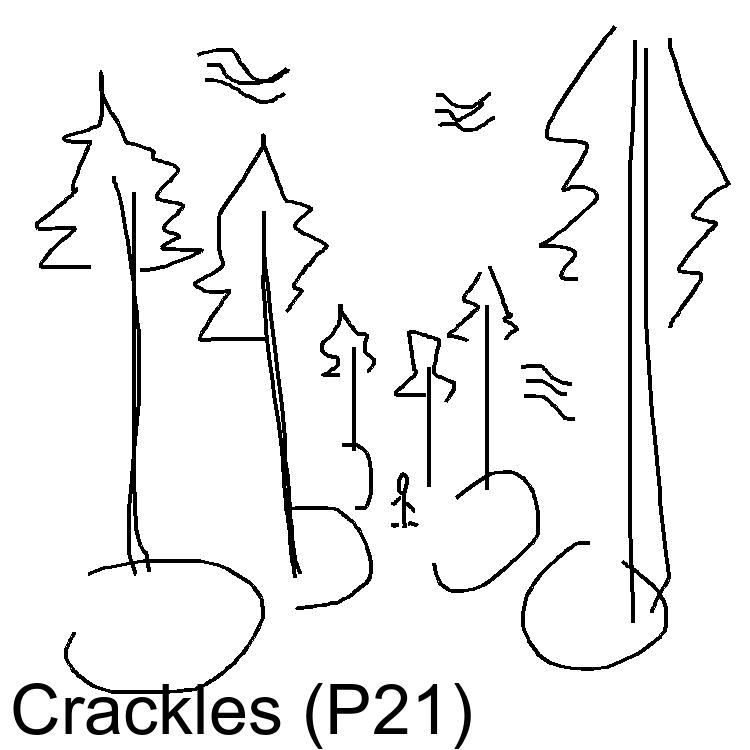} 
    & 
    \includegraphics[width=1.05\linewidth]{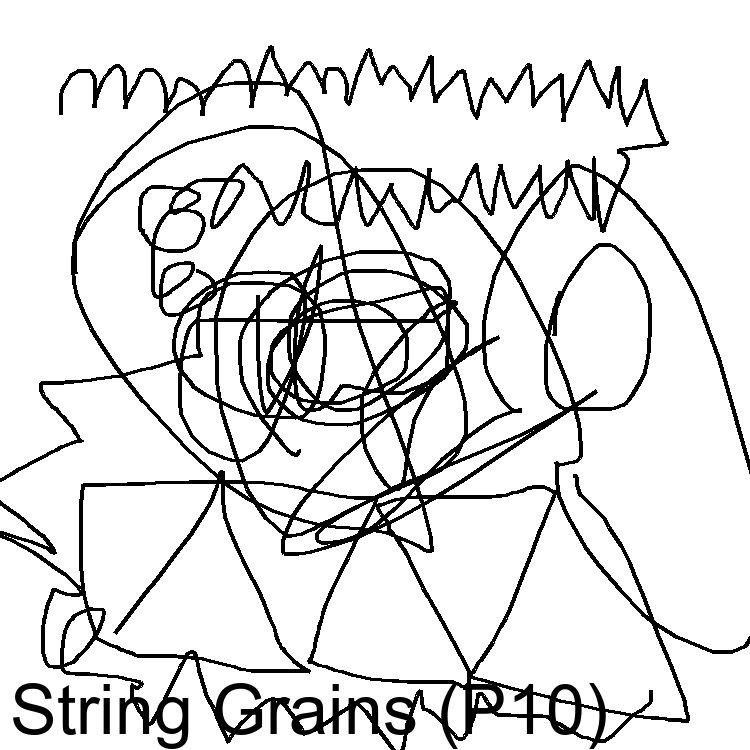} 
    & 
    \includegraphics[width=1.05\linewidth]{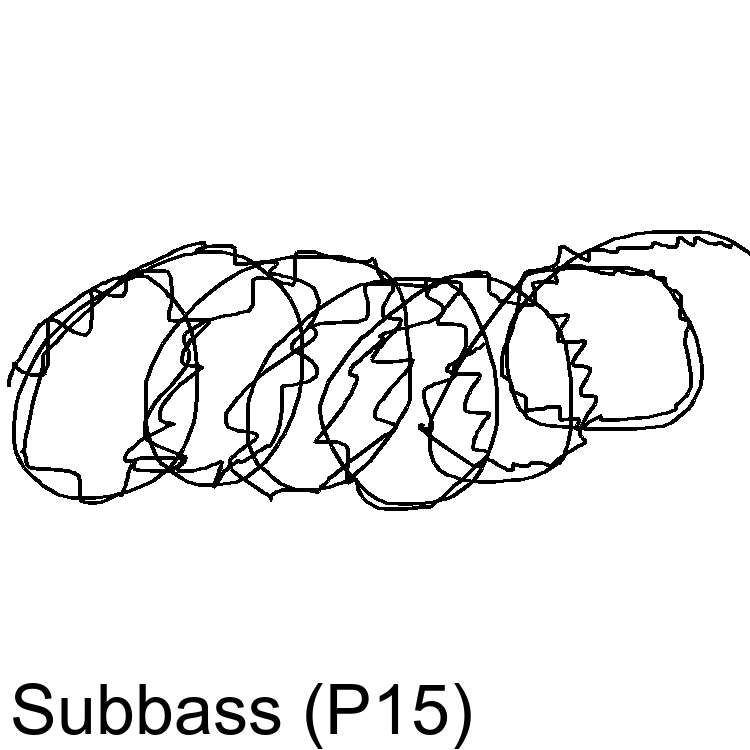} 
    &
    \includegraphics[width=1.05\linewidth]{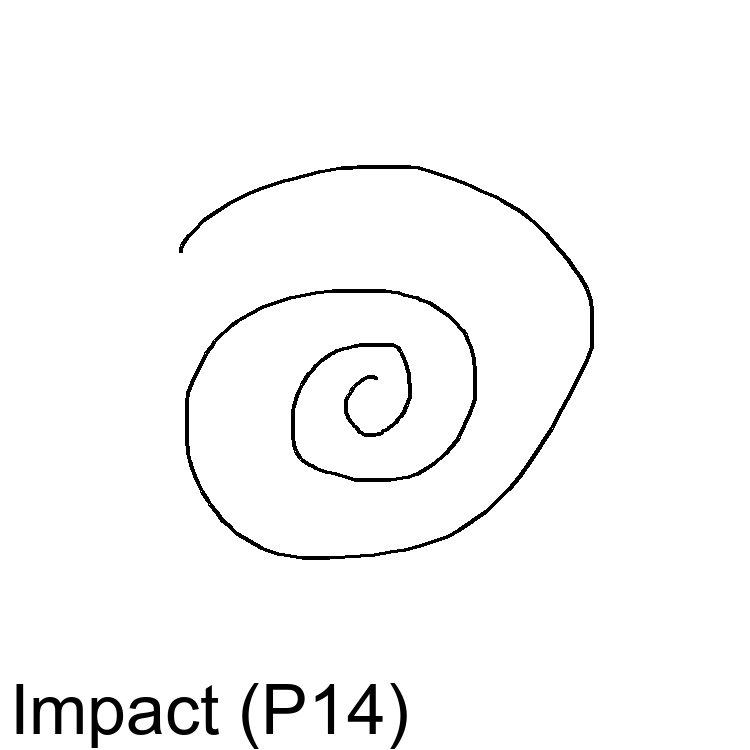}
    &
    \includegraphics[width=1.05\linewidth]{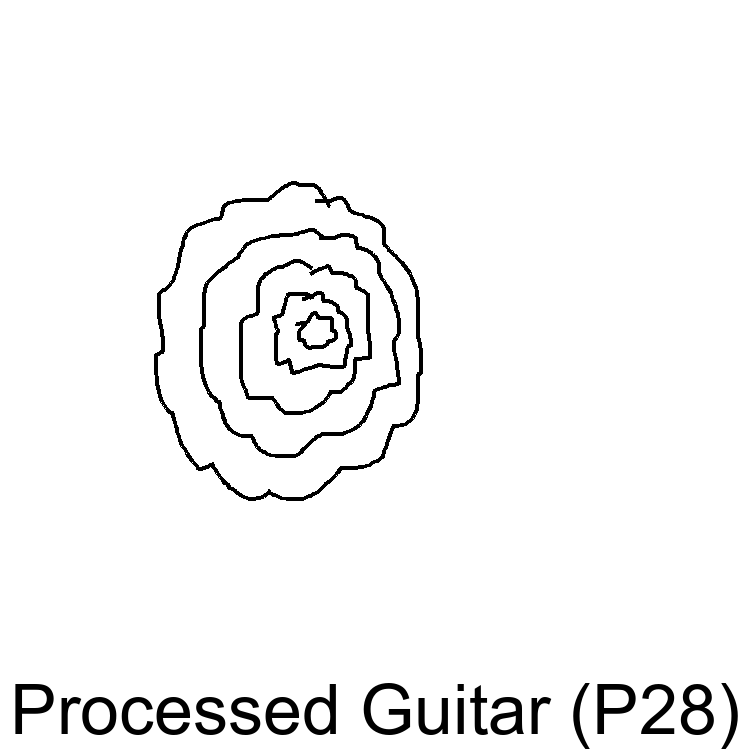}
  \\
  \hline
    \includegraphics[width=1.05\linewidth]{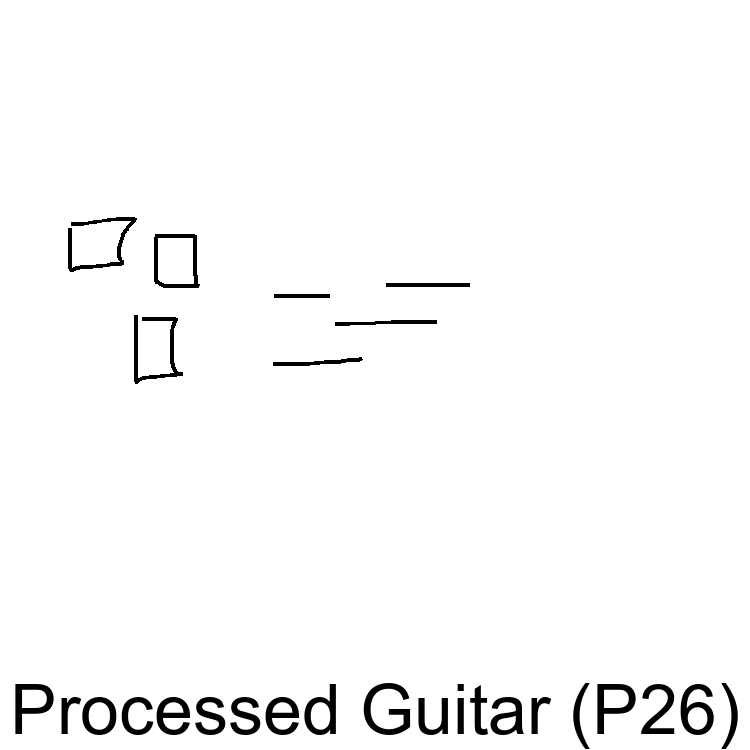} 
    & 
    \includegraphics[width=1.05\linewidth]{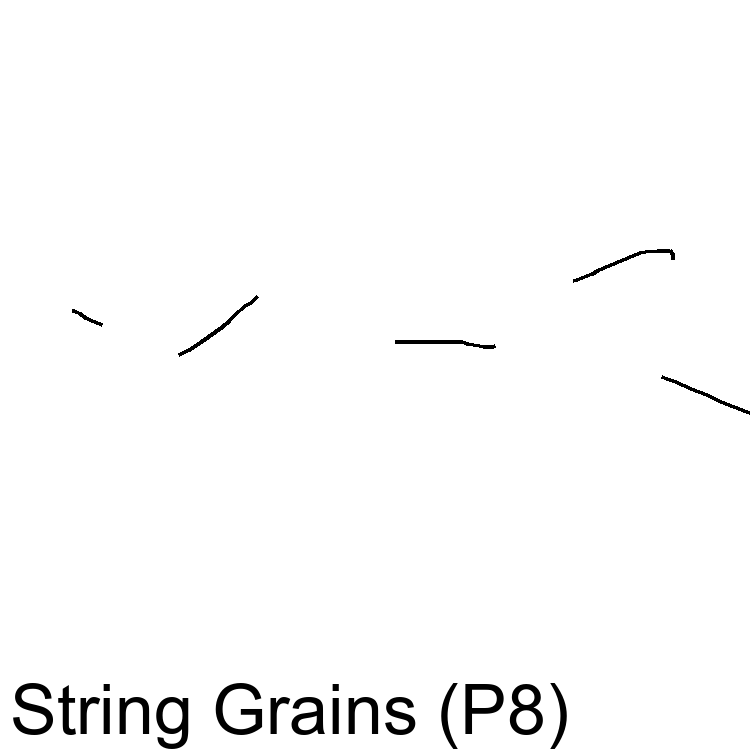} 
    & 
    \includegraphics[width=1.05\linewidth]{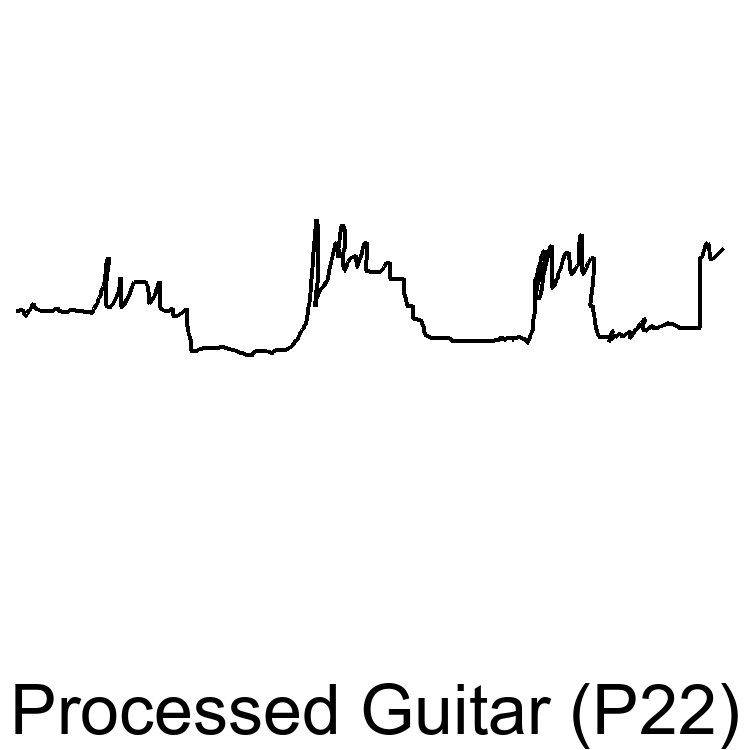} 
    & 
    \includegraphics[width=1.05\linewidth]{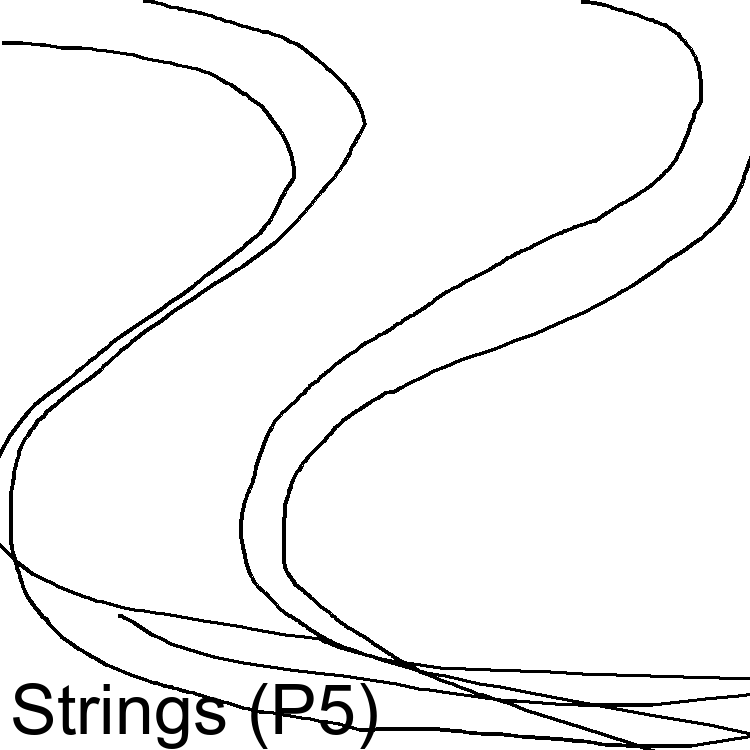}
    & 
    \includegraphics[width=1.05\linewidth]{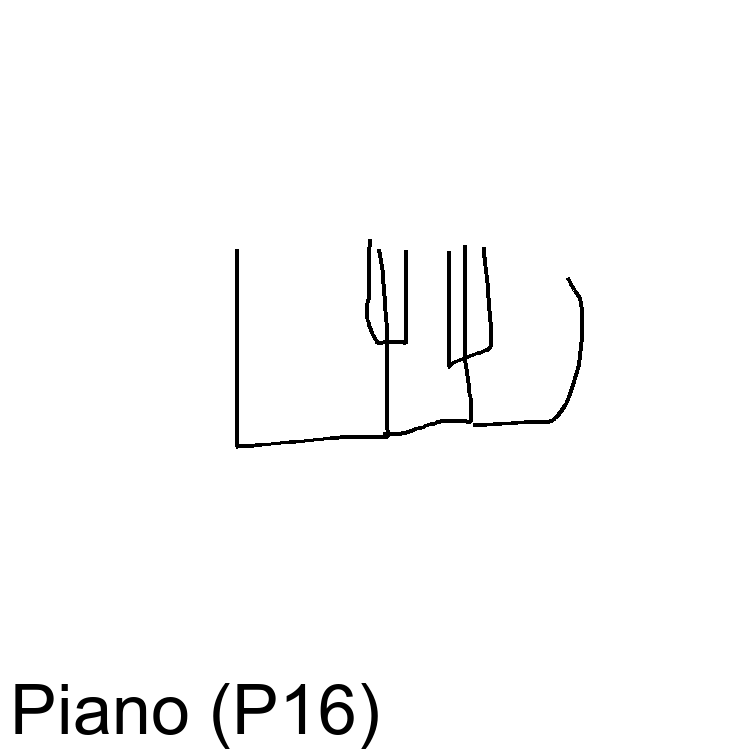} 
    & 
    \includegraphics[width=1.05\linewidth]{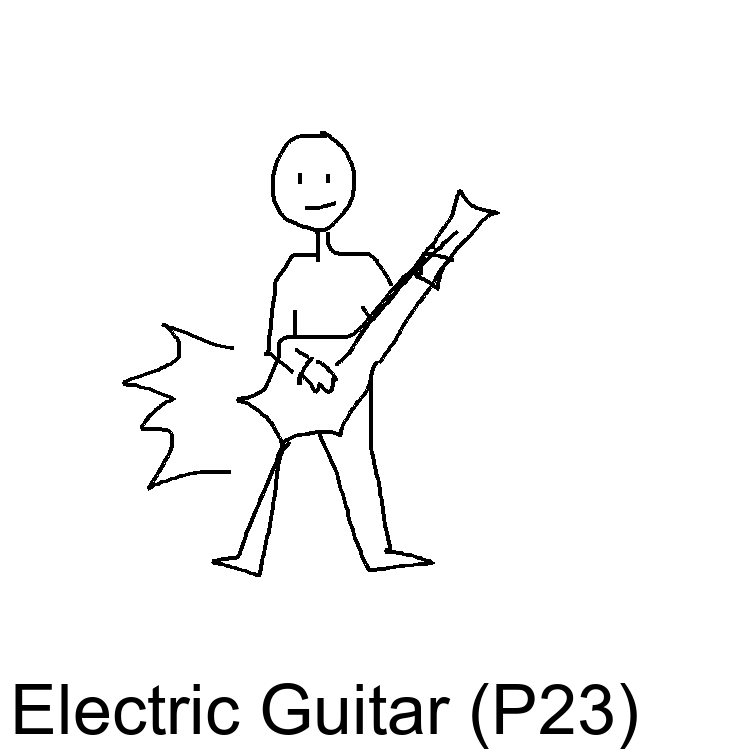} 
    & 
    \includegraphics[width=1.05\linewidth]{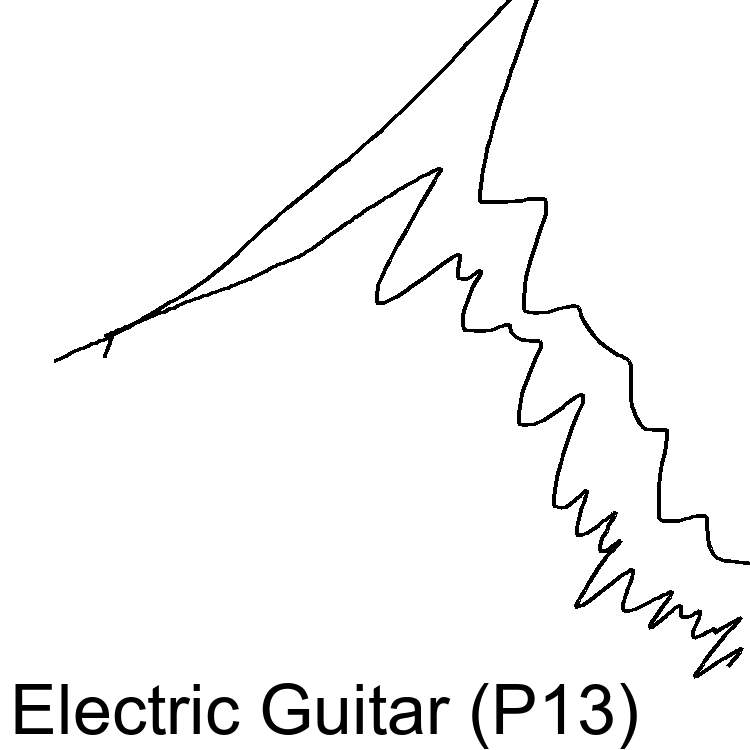} 
    & 
    \includegraphics[width=1.05\linewidth]{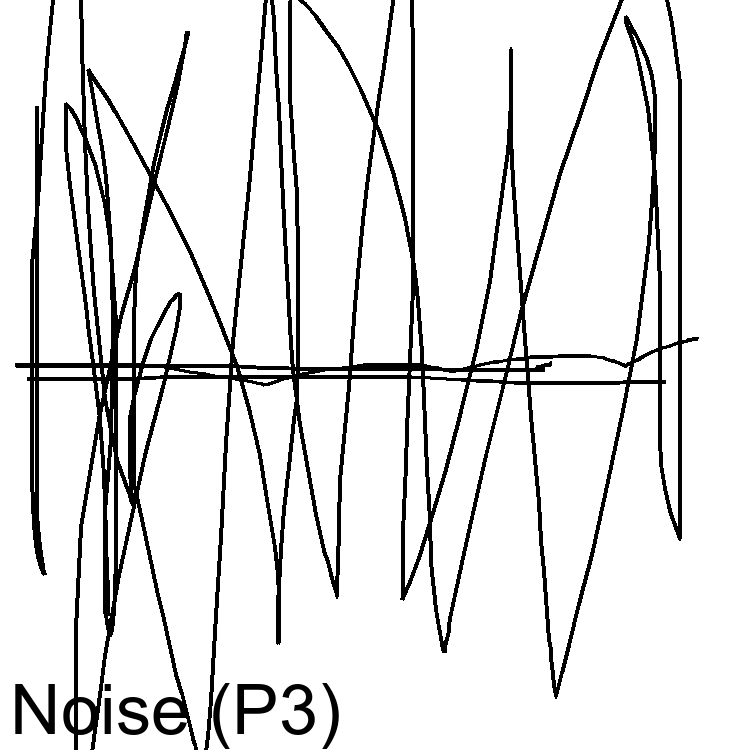} 
    & 
    \includegraphics[width=1.05\linewidth]{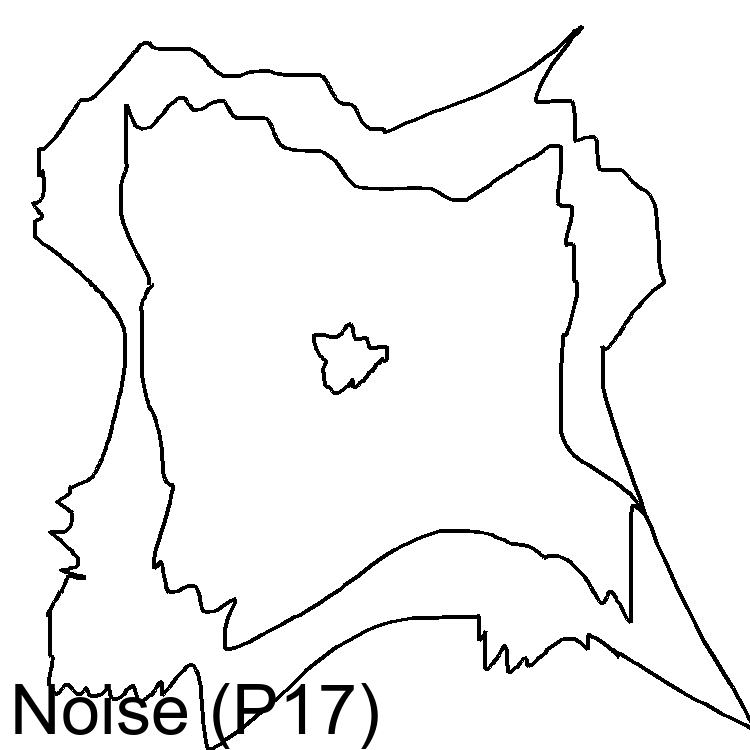}
    &
    \includegraphics[width=1.05\linewidth]{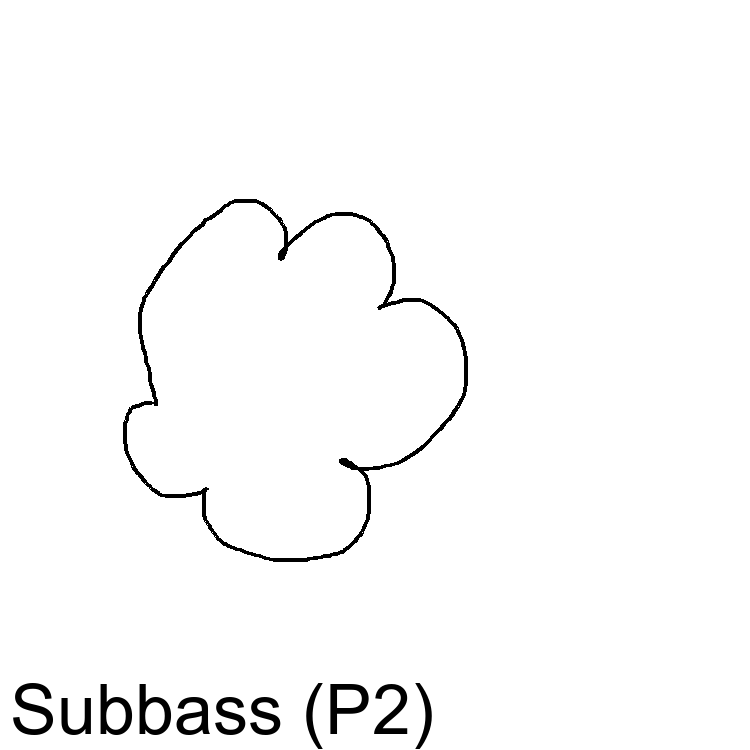}
    \\
  \hline
 \end{tabular}
\end{center}
\vspace*{-0.05in}
 \caption{ Sketch categories with examples. Category names and keywords were obtained through thematic analysis as described in Section \ref{sec:categorisation_analysis}. \textit{Objects/Scenes} mainly refers to real-world associations while other categories highlight different abstract approaches, but category clusters might overlap with a number of sketches showing characteristics of more than one category. Colours were inverted for better visibility.}
 \label{tab:sketches}
\end{table*}

\section{Analysis}
\label{sec:analysis}
This section describes the qualitative and quantitative methods used to analyse the collected data.

\subsection{Interview Analysis}
\label{sec:interview_analysis}
The interview analysis aimed to identify and summarise different approaches to the task and quantify reported difficulty. Interviews were transcribed and thematic analysis~\cite{braunUsingThematicAnalysis2006} was used to find reoccurring themes. Task difficulty was coded into hard/neutral/easy depending on the response to the question, ``How difficult did you find the task?".    

\subsection{Sketch Categorisation}
\label{sec:categorisation_analysis}
Section \ref{sec:methods} introduced the hypothesis that participants will predominately produce abstract sketches. This can be tested by dividing sketches into high-level categories that refer to their representational approach. In order to minimise bias, sketches were categorised in an open card sorting study by 6 participants (4 female, 3 musicians) who did not take part in the main study. They were asked to create a reasonable number of categories (three to ten was recommended) and write a short description for each of them. The study was completed remotely within three hours on participants' devices.\footnote{Card sorting instructions: \url{https://youtu.be/LXTlnaAciWw}} Results were reduced to two dimensions with principal component analysis (PCA) and clustered with the K-Means algorithm. The silhouette coefficient~\cite{rousseeuw1987silhouettes}, a measure of cluster goodness, was calculated to find the most suitable number of clusters between three and ten. Clusters were annotated and named with the help of keywords that were extracted from participants' descriptions using thematic analysis.   

\subsection{Sketch Feature Extraction}
\label{sec:sketch_features}
While sketch categories give an overview of the representational approaches, a more detailed, quantitative set of features is needed to compare sketches in detail and find correlations with sound characteristics using statistical analyses described in Section \ref{sec:stats}. A number of features can be calculated from the sketches' data shape and through simple arithmetic operations as demonstrated in Equations 1, 2 and 3, where \textit{$N$} is the number of strokes in a sketch and \textit{$\overline{L}$}, \textit{$\overline{T}$} and $\overline{S}$ are their average length, completion time and drawing speed. The total number of points in the \textit{$k^{th}$} stroke is described by \textit{$n_k$}. Each point has a position \textit{$x_{k_i}$} and timestamp  \textit{$t_{k_i}$}. The euclidean distance between two points is described by $d(p, q)$. 
\begin{align}
    \overline{L} & = \frac{1}{N} \sum_{k=1}^{N} \sum_{i=2}^{n_k} d\left(x_{k_{i}}, x_{k_{i-1}} \right)    
    \\
    \overline{T} & = \frac{1}{N} \sum_{k=1}^{N} t_{k_{n_k}} - t_{k_1}
    \\
    \overline{S} & = \frac{1}{N} \sum_{k=1}^{N} \sum_{i=2}^{n_k} d\left(x_{k_{i}}, x_{k_{i-1}} \right) \frac{1}{t_{k_{n_k}} - t_{k_1}}
\end{align}

Sound-shape associations are usually reported with respect to a shape's contour focusing on their ``jaggedness" or ``roundness"~\cite{adeliAudiovisualCorrespondenceMusical2014,grillVisualizationPerceptualQualities2012}. These attributes were quantified by extracting corner points divided into obtuse, right and acute angles~\cite{wolinShortStrawSimpleEffective} and curve points divided into wide and narrow shape~\cite{xiongRevisitingShortStrawImproving2009}. A qualitative review suggested that sketches differ by the number of stroke intersections that can be interpreted as the ``noisiness" of a sketch. The number of intersections was determined using an adaptation of Bresenham's rasterisation algorithm~\cite{bresenham1965algorithm}. Prior to extracting features, the sketch data was cleaned by removing consecutive points with the same position and merging two strokes if a starting point was within a five pixel distance to an end point. The number of intersections, corner and curve points is reported relative to the total stroke length of a sketch.\footnote{A detailed summary of audio and sketch feature extraction can be found at \url{http://doi.org/10.5281/zenodo.4764351}.}

\subsection{Audio Feature Extraction}
In order to investigate sound-shape associations through statistical analysis, the sound stimuli also have to be described with quantitative features. This was accomplished by computing the mean values of \textit{Centroid Frequency}, \textit{Spectral Flatness}, \textit{Zero Crossing} and \textit{Root Mean Square Power (RMS)}~\cite{peeters2004large} for each sound using the \textit{Librosa} library with a FFT window size of 2048 and hop length of 512. In addition, the timbre models proposed by Pearce et al.~\cite{pearceModellingTimbralHardness2019} provided quantified measures of \textit{Hardness}, \textit{Depth}, \textit{Brightness}, \textit{Roughness}, \textit{Warmth}, \textit{Sharpness} and \textit{Boominess}. The additional feature \textit{RMS Slope}, describing how continuous or intersected a sound is, was quantified by the slope between prominent extrema in the RMS envelope.

\subsection{Statistical Analyses}
\label{sec:stats}
Differences in sketch category counts between participant groups and between sounds were computed using Pearson's Chi-squared test and Cochran's Q test respectively. Spearman's rank coefficient was used to find significant correlations between audio features and mean sketch features between all participants. To determine whether inter-rater reliability of sound descriptions can be measured with the sketch features introduced in Section \ref{sec:sketch_features}, the ICC(2,k) model\footnote{For the ICC, sound stimuli were defined as subjects and sketch features as measurements. Repeated sounds were considered separate subjects.} of the intraclass correlation coefficient (ICC)~\cite{koo2016guideline} was deployed. Sketch features were first log-transformed to meet the normal distribution assumption of the ICC. 

\section{Results}\label{sec:results}
This section presents the results of the interview analysis, sketch categorisation, inter-rater reliability testing and correlation between audio and sketch features.

\subsection{Interview}
\label{sec:interview}
Task difficulty was reported as easy/hard/neutral by 14/8/6 participants. Participants who felt that the task was easy thought that ``there was no right or wrong'' (P4), ``it was just about being creative'' (P15) and they did not have to ``achieve something" (P10). On the contrary, others found it difficult to ``think of sound in a very visual way'' (P8) (P16). While some participants approached the task as an intuitive, creative activity, others were concerned with establishing a consistent visual language. Difficulties arose while deciding which sound characteristics to follow because ``there are too many things to consider'' like ``brightness or aggressiveness or how it [timbre] develops over time'' (P6). A consistent approach was difficult to maintain because of a ``great variety in the sounds'' (P2). Some participants reported that ``complicated ones [sounds] sounded like pictures, and then the simple ones [...] like piano notes were a lot harder to draw'' (P8) possibly because they ``hear [them] all the time'' (P1), while other participants thought that ``it’s pretty straightforward because I know a piano note more than others'' (P5).

%  One participant who did not deviate from their initial concept reported that they found themselves ``going round in circles, and question how valid the whole approach is'' (P9). 

\subsection{Sketch Categories}
\label{sec:categories}

\begin{figure}[h]
\centering
\vspace*{-0.15in}
\includegraphics[width=0.9\columnwidth]{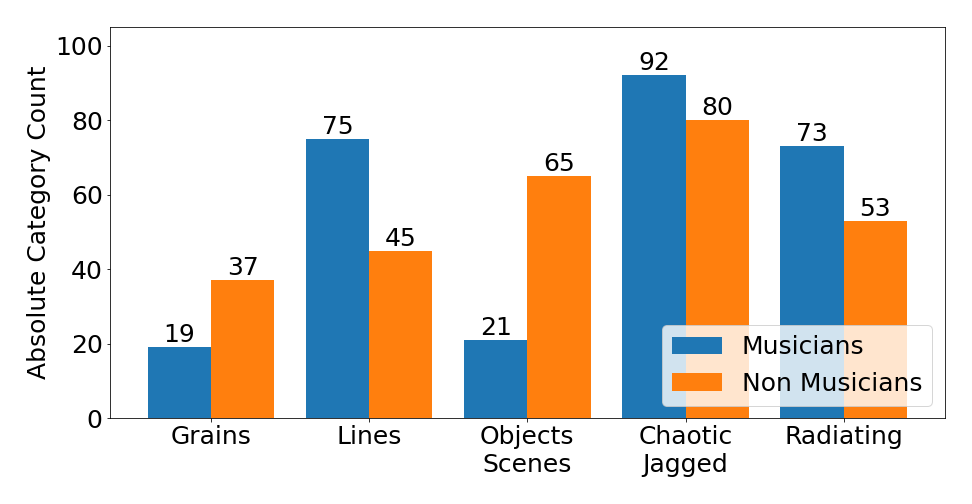}
\vspace*{-0.1in}
\caption{Absolute category counts by music proficiency \label{fig:Cat_Music}}
\end{figure}

\begin{figure}[h]
\centering
\vspace*{-0.25in}
\includegraphics[width=0.9\columnwidth]{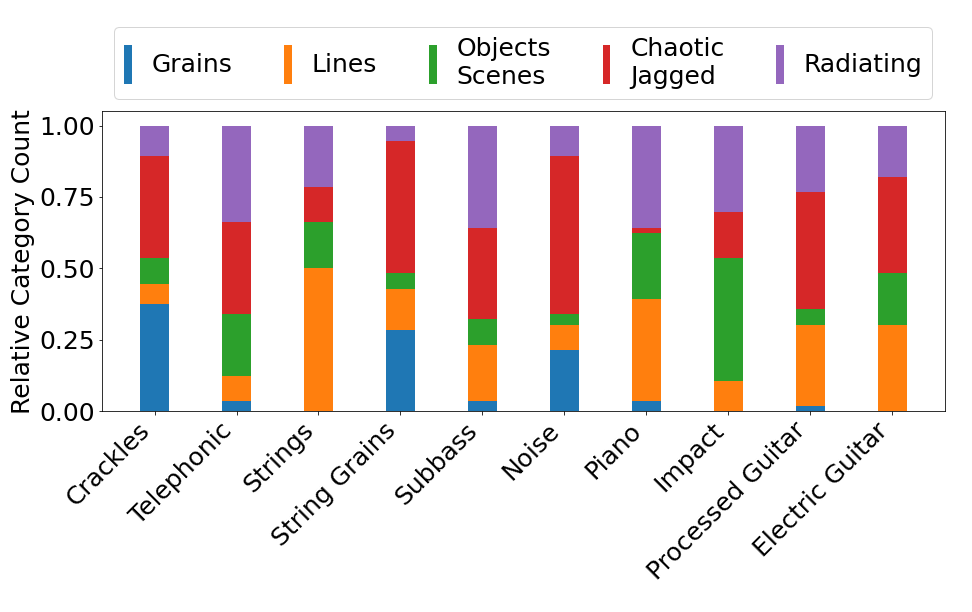}
\vspace*{-0.15in}
\caption{Relative category counts by sound stimulus\label{fig:Cat_Sound}}
\end{figure}

Analysis of the card sorting study described in Section \ref{sec:categorisation_analysis} returned an optimal number of five categories that were named: \textit{Chaotic/Jagged} (172 sketches), \textit{Radiating} (126), \textit{Lines} (120), \textit{Objects/Scenes} (86) and \textit{Grains} (56). Descriptive keywords and sketch examples for each category can be found in Table \ref{tab:sketches}. A maximal silhouette coefficient of 0.49 suggests that categories are distinguishable, but not clearly separated which is also reflected by occasionally overlapping keywords. Chi-squared test suggests that non-musicians produce \textit{Objects/Scenes} sketches more often (${\chi}^2$(1,N$=$28)$=$22.51 p$<$.0001) while musicians produce \textit{Lines} sketches at a higher rate (${\chi}^2$(1,N$=$28)$=$7.5 p$<$.01) possibly because this category contains sketches that appear to reference audio visualisations like envelopes or waveforms. Category counts for \textit{Objects/Scenes} sketches significantly differ between sounds (${\chi}^2$(9)=67.07 p$<$.0001) with post-hoc analysis revealing that \textit{Piano} and \textit{Impact} show significantly higher counts than \textit{Noise}, \textit{String Grains} and \textit{Processed Guitar} (p$<$.01 for each pair).

\subsection{Inter-rater Reliability}
\label{sec:inter-rater}

\begin{figure}[h]
\centering
\includegraphics[width=0.9\columnwidth]{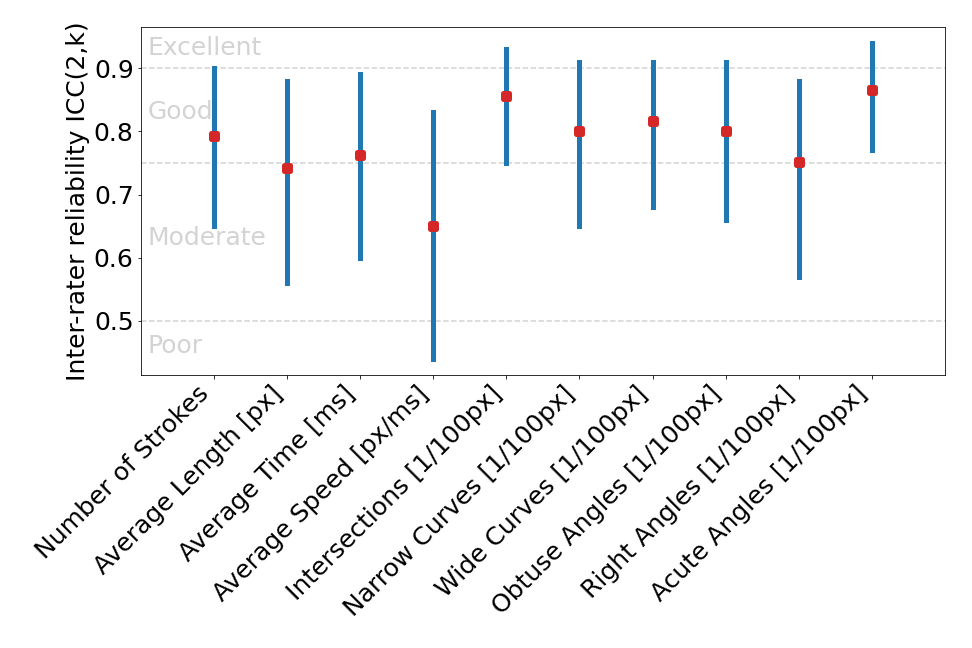}
\caption{Mean values and 95\% CI of ICC(2,k) inter-rater reliabilities for each sketch feature with evaluation guidelines proposed by Koo and Li \cite{koo2016guideline} (df1$=$19, df2$=$513, p$<$.01 for all features)\label{fig:ICC}}
\end{figure}

As seen in Figure \ref{fig:ICC}, reliability measures were good to excellent for \textit{Intersections} and \textit{Acute Angels}, poor to good for \textit{Average Speed} and moderate to good for all remaining features within the 95\% confidence interval (CI) suggesting that some level of agreement exists between participants on how to represent sounds visually and that it can be measured with the sketch features introduced in Section~\ref{sec:sketch_features}.  

\subsection{Feature Correlations}
\label{sec:correlations}

\begin{figure}[h]
\centering
\vspace*{-0.15in}
\includegraphics[width=1\columnwidth]{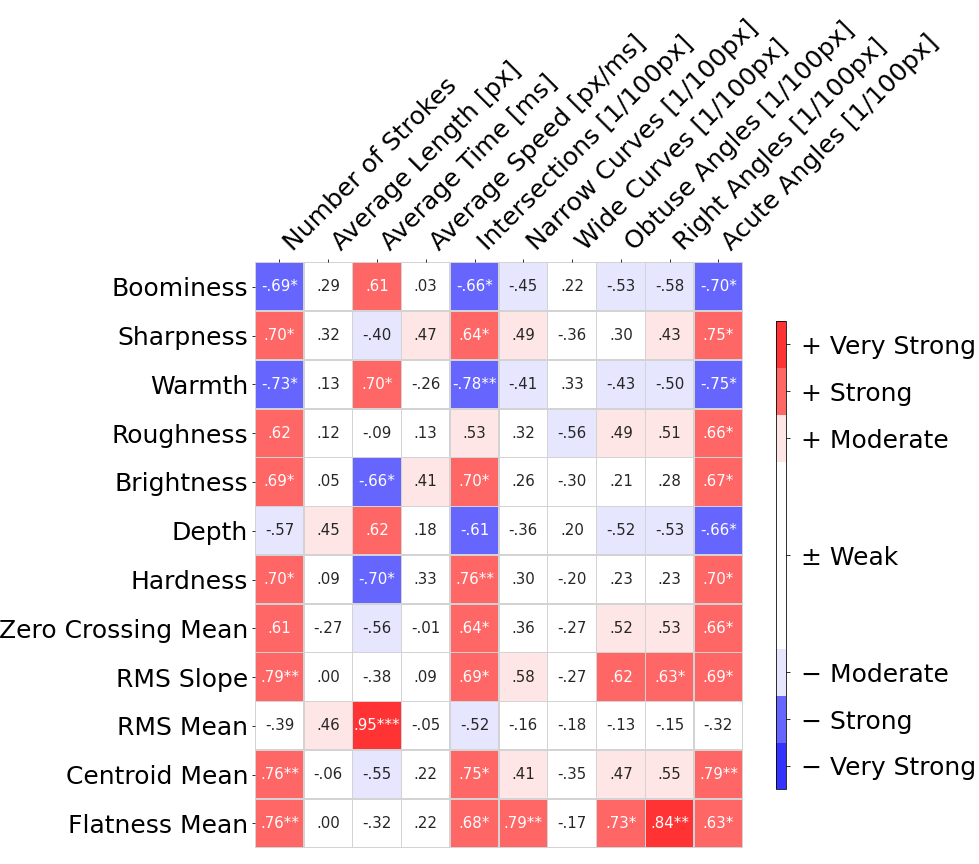}
\vspace*{-0.15in}
\caption{Spearman's rank correlation coefficients between sketch and audio features with annotated p-values: p$<$.05 (*), .01 (**), .001 (***) \label{fig:Spearman}}
\end{figure}

Several significant correlations were found between sketch and audio features. \textit{Acute Angles} (11), \textit{Intersections} (9) and \textit{Number of Strokes} (8) show the highest statistically significant (p$<$.05) number of strong (r$>$.6) and very strong (r$>$.8) correlations with audio features. The strongest correlation overall was found between \textit{RMS Mean} and \textit{Average Time} (r$=$.95, p$<$.001). Opposing audio features like \textit{Warmth} and \textit{Sharpness} showed similar absolute correlation values but opposite directions for \textit{Number of Strokes}, \textit{Intersections} and \textit{Acute Angles}.

\section{Discussion}
\label{sec:discussion}
All participants completed the study successfully, but the exploratory study design described in Section \ref{sec:procedure} benefited participants who approached the task intuitively and made it more difficult for those who aimed to follow a consistent, systematic approach. The card sorting results provide meaningful high-level categories that should, however, not be interpreted as mutually exclusive with many sketches showing elements of more than one category. Overall, participants predominately chose abstract elements like shapes and contours over scenes or icons, but this does not necessarily capture a participant's intention. For example, a sketch showing a single line might visualise the simplicity of a sound, but could also refer to its amplitude envelope. Music proficiency and the type of sound had an influence on the representational approach with depictions of real-life associations being more prevalent among non-musicians and for instrumental or environmental sounds as reported in Section \ref{sec:categories}. Some participants reported having experience with digital drawing applications and, while this was not quantified in the study, it does appear to have had an effect on their approach. A different interface, like a graphics tablet, might also have an impact on the results. The ICC analysis suggests that some level of agreement exists between participants on how to sketch sounds visually and that these sketches can be described reliably with the features introduced in Section \ref{sec:sketch_features}. However, the ICC(2,k) model used in the analysis considers the averaged measure of all raters and cannot provide information about the reliability of a single rater. A sketch-based synthesiser needs to work with the input of individual users one at a time and therefore rely on measurements with a high single rater reliability. In future work, the suitability of the sketch features has to be evaluated in that context. A large number of strong correlations between sketch and audio features was found that generally align with results from studies where existing visualisations were matched to sounds  \cite{adeliAudiovisualCorrespondenceMusical2014,grillVisualizationPerceptualQualities2012}. Sharp, rough and hard sounds result in sketches with more acute angles compared to warm or deep sounds. Contrary to expectation, curve points did not show any significant correlations as shown in Figure \ref{fig:Spearman}. This could either mean that warm sounds were represented with lines rather than curves, which is supported by significant negative correlations between \textit{Warmth}/\textit{Depth} and \textit{Intersections} or indicate that a shape's roundness was not represented well by the curve points. \textit{Number of Strokes} and \textit{Intersections}, features not commonly discussed in sound-shape research, produced strong correlations and should be considered in future work. A qualitative review led to the hypothesis that \textit{Objects/Scenes} sketches will not produce the same correlations as abstract sketches because they correspond to high-level associations rather than specific sound characteristics. However, subsets were too small for conclusive quantitative evaluation. Generally, a larger dataset would produce more robust results for all statistical tests.

\section{Conclusion}
\label{sec:conclusion}
In this exploratory study, a variety of possible user responses were found that could be expected when implementing a sketch-based sound synthesiser that uses a digital drawing interface. Results indicate that there is a general consensus about how to communicate timbre through visual sketches that can be quantified in a statistically meaningful way by extracting visual and audio features. These findings support the assumption that the development of a sketch-based synthesiser is feasible. However, an exploratory study design can only provide general, indicative results. Future work will have to focus on a specific set of parameters on which a cross-modal mapping paradigm can be built. Stricter instructions, a larger sample size, possibly focusing on a specific type of participant, and a smaller, targeted set of sound stimuli, for example using only synthesised pads, could be beneficial to produce more detailed results.

\begin{acknowledgments}
EPSRC and AHRC Centre for Doctoral Training in Media and Arts Technology (EP/L01632X/1).
\end{acknowledgments}

% this inserts a column-break
%\pagebreak

%%%%%%%%%%%%%%%%%%%%%%%%%%%%%%%%%%%%%%%%%%%%%%%%%%%%%%%%%%%%%%%%%%%%%%%%%%%%%
%bibliography here
\bibliography{icmc2021template}

\end{document}